\documentclass[aps,twocolumn]{revtex4}

\usepackage{graphicx}
\usepackage{amsmath}
\usepackage{amssymb}
\usepackage{hyperref}

\begin{document}
\title{
Viral evolution under the pressure of an adaptive immune system -\\
optimal mutation rates for viral escape
}
\author{Christel Kamp}
\email{kamp@theo-physik.uni-kiel.de}
\affiliation{Institut f\"ur Theoretische Physik, Universit\"at Kiel, Germany}
\author{Claus O. Wilke}
\affiliation{Digital Life
Laboratory 136-93, California Institute of Technology, Pasadena, USA}
\author{Christoph Adami}
\affiliation{Digital Life
Laboratory 136-93, California Institute of Technology, Pasadena, USA}
\affiliation{Jet Propulsion Laboratory 126-347, California Institute of
Technology, Pasadena, USA}
\author{Stefan Bornholdt}
\affiliation{Institut f\"ur Theoretische Physik und Astrophysik, Universt\"at
Kiel, Germany}
\affiliation{Interdisziplin\"ares Zentrum f\"ur Bioinformatik, Universit\"at Leipzig, Germany}

\begin{abstract}
\noindent
  Based on a recent model of evolving viruses competing with an adapting
  immune system~\cite{kamp:bornholdt:2002.1}, we study the conditions under
  which a viral quasispecies can maximize its growth rate. The range of
  mutation rates that allows viruses to thrive is limited from above due to
  genomic information deterioration, and from below by insufficient sequence
  diversity, which leads to a quick eradication of the virus by the immune
  system. The mutation rate that optimally balances these two
  requirements depends to first order on the ratio of the inverse of the
  virus' growth rate and the time the immune system needs to develop a
  specific answer to an antigen. We find that a virus is most viable if it
  generates exactly one mutation within the time it takes for the immune
  system to adapt to a new viral epitope.  Experimental viral mutation rates,
  in particular for HIV (human immunodeficiency virus), seem to suggest that
  many viruses have achieved their optimal mutation rate.
\end{abstract}

\maketitle

\section{Introduction}
Since Eigen and Schuster introduced the concept of a
quasispecies~\cite{eigen:1971,eigen:schuster:1979}, it has become a standard
model to describe molecular and viral evolution. If a simple, single-peaked
fitness landscape is assumed, quasispecies theory predicts that error-prone
replication leads to the formation of a central ``master sequence'',
surrounded by a cloud of mutant sequences.  For viral evolution, this implies
that any ``wild-type'' sequence is accompanied by a cloud of related mutants
that, as a whole, represent a target for the immune system. The quasispecies
approach to molecular evolution has been the object of detailed
investigations, often supported by techniques of statistical physics
\cite{schuster:sigmund:1983,demetrius:1983,demetrius:sigmund:1985,schuster:1986,leuthaeusser:1986,leuthaeusser:1987,schuster:swetina:1988,nowak:schuster:1989,tarazona:1992,bonhoeffer:stadler:1993,pastor-satorras:sole:2001,domingo:holland:2001}
revealing the characteristic features of such systems, including the
occurrence of an error catastrophe. The latter characterizes a system in which
a critical mutation rate exists beyond which the genomic information is
irretrievably lost to mutations, i.e., beyond which selection ceases to
operate~\cite{nowak:schuster:1989,tarazona:1992,bonhoeffer:stadler:1993,alves:fontanari:1997,campos:fontanari:1998,campos:fontanari:1999,altmeyer:mccaskill:2001}
(for an in-depth discussion of error catastrophes and related phenomena see
also \cite{hermisson:baake:2002}). The destabilizing effect of increased
mutation rates has been observed for various viruses, including
HIV~\cite{loeb:mullins:1999} and Poliovirus~\cite{crotty:andino:2001}.
\newline
Recently, various extensions of the Eigen-Schuster model have been considered,
in particular involving the shape of the fitness peaks and the landscapes'
time-dependence. While the shape of the fitness function influences the
robustness of a species to
mutations~\cite{vannimwegen:huynen:1999,wilke:adami:2001,wilke:2001}, a
behavior qualitatively different from the standard results can be observed for
{\it non-stationary} fitness
landscapes~\cite{nilsson:snoad:2000,wilke:martinetz:2001}.  In rapidly
changing environments, a second catastrophe emerges besides the well-known
error catastrophe, termed ``adaptation catastrophe''. In a changing
environment, sequence replication must occur with a non-vanishing error rate
to enable the species to keep up with the environmental changes. (In static
landscapes, a zero mutation rate is ultimately optimal because it maximizes
the average global fitness of the population.) Indeed, a selective advantage
for so called ``mutator mutants'' (or ``general
mutators''~\cite{devisser2002}) has been observed for {\em Escherichia coli}
and {\em Salmonella enterica} under challenging living conditions
\cite{leclerc:cebula:1996,sniegowski:lenski::1997,giraud:taddei:2001}.
\newline
For viruses in the environment of an adaptive immune system, the fitness landscapes
for both the virus and the immune system are dynamically generated by a
co-evolutionary process. This dynamics can be studied within the quasispecies'
framework if the quasispecies character of {\em both} the viral population and the
motifs of immune receptors is acknowledged. In an immune response, the presence of
an antigenic epitope induces the proliferation of the corresponding immune receptor
sequence. This ``master'' sequence is associated with a cloud of closely related
receptor sequences that emerge from somatic hypermutation of B-cells in the
germinal centers~\cite{harris:maizels:1999}. Competition between a viral population
and an adaptive immune system takes place via an asymmetric coupling: while the
immune quasispecies is strongly attracted by the virus, the viral quasispecies is
driven away from its current master sequence by the immune system. This
predator-prey-dynamics results in a migration through sequence space as observed in
many infectious diseases, such as
HIV~\cite{ganeshan:wolinsky:1997,allen:watkins:2000}.

The co-evolutionary dynamics within an infected host was recently formalized within
a model relying only on a few dynamical rules~\cite{kamp:bornholdt:2002.1},
recapitulated in the following section. Here, we focus on the implications of an
optimal immune response within this framework, and consider the conditions that
correspond to maximal \textit{viral} fitness. Finally, we compare known viral
mutation rates to those expected if a viral population has achieved an
optimal mutation rate.

\section{Virus--immune system co-evolution}
\begin{figure}[h]
\centerline{
  \includegraphics[width=7cm]{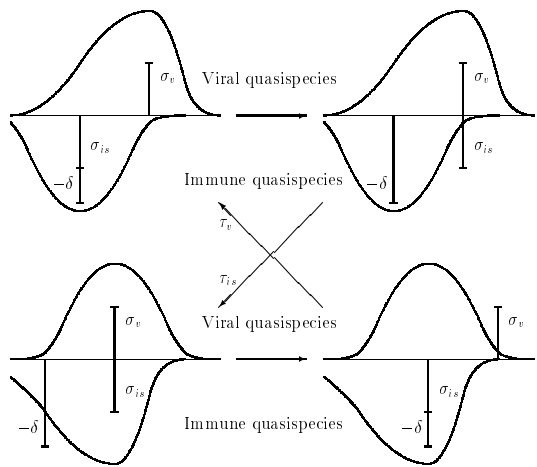}
}
\caption{\label{fig:dyn}Co-evolution of viral and immune quasispecies.}
\end{figure}

Let us assume that the viral and the immunological quasispecies alike experience a
single-peaked fitness function (Figure \ref{fig:dyn}), albeit one that can change
in time. Let us assume further that at any particular time, the (viral) master
sequence of length $n$ grows at a rate $\sigma_v$ (much larger than the ``off-peak''
or background-fitness $\eta_v$), and similarly for the immune system:
$\sigma_{is}\gg\eta_{is}$. Such a simple immunological fitness function results
from a reduction of the viral impact to induce proliferation of immune cells to its
master sequence. Analogously, only the dominant immune sequence imposes a decay
rate $\delta$ on its complementary viral sequence. Both species replicate
imperfectly, with copy fidelities $q_v<1$ and $q_{is}<1$ (denoting the probability
for correct duplication of a monomer drawn from an alphabet of size $\lambda$). 

The
virus--immune system interaction is implemented by the following dynamic rules that
are cyclicly iterated, leading to the quasispecies' migration through sequence
space:
\begin{enumerate}
\item{Once the immune system imposes a decay rate $\delta>0$ on the
viral master sequence (centered at the viral fitness peak), the narrow niche of the
virus is assumed to move to an arbitrary sequence of the first error class}.
\item{The viral quasispecies adapts to this new fitness peak on a time scale
$\tau_v$}.
\item{The fitness peak of the immune quasispecies is adjusted, and moves to the new
maximum of the viral distribution}.
\item{The immune system adapts to the new fitness peak on the time scale
$\tau_{is}$}.
\end{enumerate}
As discussed previously~\cite{kamp:bornholdt:2002.1}, the dynamically
generated time scale $\tau = \tau_v+\tau_{is}$ can be approximated by the
two expressions
\begin{equation}
 \tau_v\approx -
\frac{\ln\left(\frac{1-q_v}{\lambda-1}\right)}
{q_v^n(\sigma_v-\eta_v)+\delta}\;
\end{equation}
and
\begin{equation}
\tau_{is}\approx -\frac{\ln\left(
\frac{1-q_{is}}{\lambda-1}\right)}{q_{is}^n
(\sigma_{is}-\eta_{is})}.
\end{equation}

The relative growth of the (moving) viral and immunological master sequences in
comparison to the environmental (background) sequences' growth can be determined
as~\cite{kamp:bornholdt:2002.1,nilsson:snoad:2000}:
\begin{equation} \label{eqn3}
\kappa_i
=\frac{\left(e^{(q_i^n\sigma_i-\eta_i)\tau}-e^{(q_i^n\eta_i-\eta_i)\tau}\right)
(1-q_i)\sigma_i}{(\lambda-1)(\sigma_i-\eta_i)q_i},\quad i\in\{v,is\},
\end{equation}
leading to the conditions
\begin{equation}
\kappa_v>1,\qquad\kappa_{is}>1
\end{equation}
for viability of the viral and immunological species, respectively. The regimes of
(co-)existence of the two quasispecies can be determined by analyzing $\kappa_v$
and $\kappa_{is}$. In particular, the viral quasispecies is subject to both a
classical error catastrophe at high mutation rates, and an adaptation catastrophe
for small mutation rates. In contrast, the immune system (as the driving force) is
not subject to a limiting migration velocity, and accordingly only displays the
classical error catastrophe \cite{kamp:bornholdt:2002.1}.

\section{Optimal viral mutation rate}
Having derived the relations quantifying viral as well as immunological viability,
we can now deduce optimal strategies for both the virus and the immune system. The
immune system attempts to minimize viral growth ($\frac{\partial\kappa_v}{\partial
q_{is}}\stackrel{!}{=}0$) which implies the relation
\begin{equation}
\mu_{is}-1-n_{is}\mu_{is}\ln\left(\frac{\mu_{is}}{\lambda-1}\right)=0\;; \qquad
\mu_{is}=1-q_{is}.
\end{equation}
between the optimal immune receptor size $n_{is}$ and the per-site mutation
probability $\mu_{is}$. This prediction and how it fares against the background of
experimental data has been discussed in \cite{kamp:bornholdt:2002.1}. Below, we
extend this approach to derive the conditions for optimal viral escape from an
immune response.

\begin{figure}[h]
\centerline{
  \includegraphics[width=8cm]{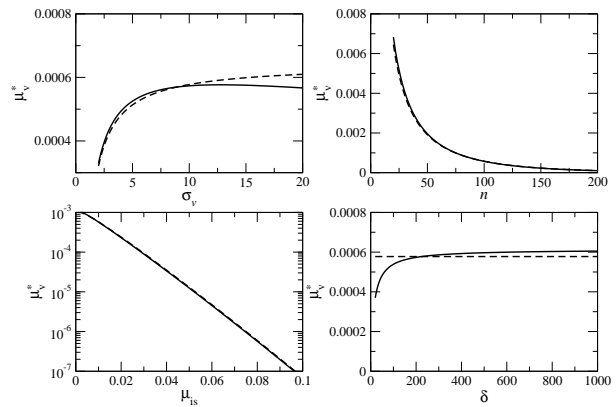}
}
\caption{\label{fig:exact-vs-approx}Optimal per-site mutation rate
  $\mu_v^\ast$,
  comparison between the analytic approximation as given by equation
  \eqref{p_v-opt} (dashed lines) and the numerical solution to equation
  \eqref{kappa-cond} (solid lines). 
  Parameters are $\sigma_v=10$, $\eta_v=1$,
  $\sigma_{is}=10$, $\eta_{is}=1$,
  $q_{is}=0.99$, $n=100$, $\delta=200$, $\lambda=4$, unless specified otherwise in
  the plot.}
\end{figure}

Let us first approximate $\kappa_v$ in Eq.~(\ref{eqn3}) by
\begin{equation}\label{kappa-approx}
 \kappa_v \approx \frac{1-q_v}{\lambda-1}\exp[(q_v^n\sigma_v-\eta_v)\tau]\,,
\end{equation}
using $\sigma_v\gg\eta_v$, $q_v\approx 1$. Optimizing viral viability conditions is
akin to maximizing the viral species' relative growth rate $\kappa_v$ such that
\begin{equation}\label{kappa-cond}
  \frac{\partial \kappa_v}{\partial q_v} = 0\;.
\end{equation}
Inserting $\tau=\tau_v+\tau_{is}$ into \eqref{kappa-approx} leads 
the equivalent condition
 \begin{eqnarray}\label{kappa-derivative}
  0 &=&\big(q_v^n(\sigma_v-\eta_v)+\delta\big)
  \big(n(q_v-1)q_v^{2n}\sigma_v^2\tau_{is}\notag\\
   &+&\delta[q_v+(q_v-1)nq_v^n\sigma_v\tau_{is}]\notag\\
  &+&\eta_v[q_v-q_v^{n+1}-(q_v-1)nq_v^{2n}\sigma_v\tau_{is}]\big)\notag\\
   &+& nq_v^n(q_v-1)(\eta_v^2-\delta\sigma_v-\eta_v\sigma_v)
    \ln\Big(\frac{1-q_v}{\lambda-1}\Big)\;.
\end{eqnarray}
We can simplify this expression in the following manner. Writing
(\ref{kappa-derivative}) in terms of the mutation probability $\mu_v=1-q_v$ rather
than the copy-fidelity $q_v$ allows us to expand (\ref{kappa-derivative}) in terms
of $\mu_v$ (while leaving the term in $\ln \mu_v$ untouched).
Assuming furthermore that $\delta\gg \sigma_v\gg \eta_v$, and $n\gg 1$, we find
\begin{equation}
  \frac{\partial \kappa_v}{\partial q_v} = 0 \quad
\Leftrightarrow\quad \delta^2+n\delta \sigma_v(\ln \mu_v-\delta\tau_{is})\mu_v\approx0\,.
\end{equation}
We now proceed to determining the root of this expression. While this can be done
numerically (see below), we first attempt to obtain an analytical approximation
that permits an intuitive interpretation. For this purpose, it is allowable to
assume $\ln \mu_v\approx const$, as $\ln \mu_v$ is a slowly varying function of
$\mu_v$. The optimal per-site mutation probability $\mu_v^\ast$ then follows as
\begin{equation}\label{p_v-opt}
  \mu_v^\ast = \frac{1}{n\sigma_v(\tau_{is}-const/\delta)}
    \approx \frac{1}{n\sigma_v\tau_{is}}\,.
\end{equation}
Figure \ref{fig:exact-vs-approx} shows a comparison between the optimal mutation
rate $\mu_v^\ast$ as given by the approximation~\eqref{p_v-opt}, and the exact
solution $\mu_v^\ast$ obtained numerically from \eqref{kappa-derivative}. Despite
the many approximations that have entered the derivation of \eqref{p_v-opt}, the
analytic approximation is in good agreement with the numerical results.
Improvements to the analytic approximation are possible if we neglect fewer of the
higher order terms 
\cite{footnote}.

\begin{figure}[h]
\centerline{
  \includegraphics[width=4cm]{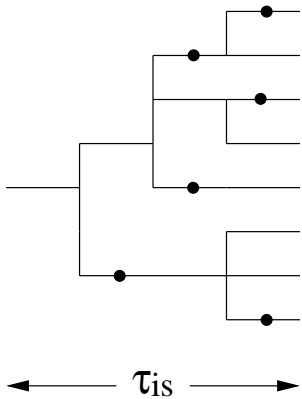}
}
\caption{\label{fig:opt-geneal}Regrowth from a single virus particle to a
  population size of eight, within the time-span $\tau_{is}$ (dots indicate
  mutations). The virus can best evade the immune system if almost every
  virion in the population at $t=\tau_{is}$ differs from the initial virion by
  exactly one mutation.}
\end{figure}

Let us now rewrite \eqref{p_v-opt} in terms of the optimal \textit{genomic}
mutation rate
\begin{equation}\label{mug-opt}
\mu_v^{G\ast}:=n\mu_v^\ast =\frac{1}{\sigma_v\tau_{is}}\;.
\end{equation}
This form suggests the following intuitive interpretation. The immune system adapts
to a new virus strain within a time-span $\tau_{is}$, while the virus replicates in
a time-span $1/\sigma_v$. The ratio between these two time scales measures the
duration of one generation of the virus in units of the response time of the immune
system. Hence, Equation \eqref{mug-opt} implies that the virus can optimally evade
the immune system if the virus suffers on average \textit{one} mutation per genome
within the time the immune system needs to adapt to a new strain
(Fig.~\ref{fig:opt-geneal}).  

This condition guarantees that a maximal number of
virions have mutated away from the epitope to populate its first error class,
\textit{precisely} at that point in time when the immune system has adapted to
attack the new viral quasispecies.

If a viral quasispecies optimizes its mutation rate according to
Eq.~(\ref{mug-opt}), we expect to see this reflected in a relation between the
mutation rate and genome size, such that their product is constant (given a
particular generation time $1/\sigma_v$). Optimization of genomic mutation rate can
take place via an optimization of sequence length, given any particular per-site
mutation rate. Table \ref{mu-n-tab} shows that the genomic mutation rate $\mu_v^G$
only slightly varies within the class of RNA viruses, which presumably have a
similar generation time. This is well in agreement with the prediction
\eqref{mug-opt}.

\begingroup
\squeezetable
\begin{table}[h!]
\begin{center}
\begin{tabular}{|l|c|c|c|}
\hline
\footnotesize{Organism} & $n$ & $\mu_v$ & $\mu_v^G=n\mu_v$ \\
\hline
\multicolumn{4}{|l|}{\footnotesize{Lytic RNA-based viruses \protect\cite{drake:1993}}}\\
\hline
\footnotesize{Poliovirus} & $7.4\cdot 10^3$ & $1.1\cdot 10^{-4}$ & $0.81$\\
\hline
\footnotesize{Influenza A Virus} & $13.6\cdot 10^3$ &  $>7.3\cdot 10^{-5}$ & $0.99$\\
\hline
\multicolumn{4}{|l|}{\footnotesize{RNA-based Retroviruses \protect\cite{drake:crow:1998,drake:1993}}}\\
\hline
\footnotesize{Spleen Necrosis Virus} & $7.8\cdot 10^3$ & $2.0\cdot 10^{-5}$ & $0.16$\\
\hline
\footnotesize{Molony Murine Leukemia Virus}& $8.4\cdot 10^3$ & $>3.5\cdot 10^{-6}$ & $0.029$\\
\hline
\footnotesize{Rous Sacroma Virus} & $9.3\cdot 10^3$ & $4.6\cdot 10^{-5}$ & $0.43$\\
\hline
\footnotesize{HIV-1} & $9.2\cdot 10^3$ & $2.4\cdot 10^{-5}$ & $0.22$\\
\hline
\end{tabular}
\caption{\label{mu-n-tab}Genomic length $n$ and spontaneous mutation rates per base
pair and replication $\mu_v$ for RNA-based viruses that compete with advanced
immune systems, as well as genomic mutation rate $\mu_v^G=n\mu_v$. Note that this
product is an approximation for $\mu_v^G=1-(1-\mu_v)^n$ for $n\mu_v<1$. Data are
reproduced from \protect\cite{drake:crow:1998,drake:1993,drake:holland:1999}.}
\end{center}
\end{table}
\endgroup

Given the adaptation time of the immune system $\tau_{is}$ and the generation time
$1/\sigma_v$, we can test the prediction Eq.~(\ref{mug-opt}) more specifically. The
adaptation time $\tau_{is}$ is the time necessary for the immune system to develop
a specific answer to an antigen. For most systems, this can be estimated to take
between 7 to 14 days \cite{roitt}. The generation times of viral species of course
vary, but data from HIV-1 is available. 

\begingroup
\squeezetable
\begin{table}[h!]
\begin{center}
\begin{tabular}{|c|c|c|c|c|}
\hline
 & $\sigma_v$ $[d^{-1}]$ & $\tau_{is}$ $[d]$ & $(\sigma_v\tau_{is})^{-1}$ & $\mu^G_v$\\
\hline
\footnotesize{HIV-1} &
$0.4...3.5$ & $7...14$ &  $0.02...0.36$ & $0.22$\\
\hline
\end{tabular}
\end{center}
\caption{\label{shiv-tab}Comparison of the genomic mutation rate $\mu^G_v$
of HIV-1 with the theoretical estimate $(\sigma_v\tau_{is})^{-1}$
from formula \eqref{mug-opt}.
Data are reproduced from
\protect\cite{little:havlir:1999,perelson:ho:1996}}
\end{table}
\endgroup

Table \ref{shiv-tab} shows that the optimal
genomic mutation rate as predicted by formula \eqref{mug-opt} is well within the
range of the experimentally determined rate. This suggests that HIV-1 has adapted
its mutation rate to optimally escape the immune system as well as the error
catastrophe.
\section{Summary}
The dynamics of co-evolution between virus and  immune system can be studied
within the framework of molecular evolution in time-dependent fitness landscapes,
in which a constantly changing, polymorphic, viral population competes with an
immune system adapting to keep track of the viral changes. Such an analysis~\cite{kamp:bornholdt:2002.1} reveals
an optimal mutation rate for the immune system (so as
to constrain the range of mutation rates within which the virus is stable) that
appears to be compatible with available data. The same
formalism can be used to determine the optimal \textit{viral} mutation rate, by
maximizing the speed of adaptation while minimizing information loss due to
mutations. It follows that the optimal viral mutation rate is reached if a sequence
undergoes on average \textit{one} mutation within the time it takes for the immune
system to adapt to the viral genomic signature, thus barely staying ahead of the
immune system. Such optimal mutation rates are compatible with experimentally
determined ones, and suggest that the constancy of genomic mutation rates within
viral classes (while sequence length and per-site mutation rates vary over many
orders of magnitude) can be explained by selection favoring viral strains at or
near the optimal rate.

\vskip 0.5cm \noindent{\large \bf  Acknowledgements} \vskip 0.25cm \noindent This
research was supported in part by the National Science Foundation under Contract
No. DEB-9981397. Part of this work was carried out at the Jet Propulsion Laboratory
under a contract with the National Aeronautics and Space Administration.
Finally, C. Kamp would like to thank the Stiftung der Deutschen Wirtschaft
for financial support.


\begin{thebibliography}{21}
\bibitem[41]{footnote}
A significant improvement for small $\delta$ can be
obtained if instead of completely neglecting the logarithmic term, we replace it
with a constant (e.g., $\ln \mu_v\approx -7$ for $\mu_v$ between about $10^{-4}$
and $10^{-2}$)
\end{thebibliography}
\end{document}